\documentclass[conference]{IEEEtran}
\IEEEoverridecommandlockouts
\newlength \figwidth
\usepackage{cite}
\usepackage{multirow}
\usepackage{amsmath,amssymb,amsfonts}
\usepackage{algorithmic}
\usepackage{graphicx}
\usepackage{textcomp}
\usepackage{xcolor}
\usepackage{float}
\usepackage{tabulary}
\usepackage{dblfloatfix}
\usepackage[multiple]{footmisc}
\usepackage{enumitem}
\def\BibTeX{{\rm B\kern-.05em{\sc i\kern-.025em b}\kern-.08em
    T\kern-.1667em\lower.7ex\hbox{E}\kern-.125emX}}
    
\newfont{\bbb}{msbm10 scaled 500}

\newfont{\bb}{msbm10 scaled 1100}

\newcommand{\ds}{\displaystyle}

\newcommand{\executeiffilenewer}[3]{%
\ifnum\pdfstrcmp{\pdffilemoddate{#1}}%
{\pdffilemoddate{#2}}>0%
{\immediate\write18{#3}}\fi%
}

\begin{document}

\bstctlcite{IEEE_nodash:BSTcontrol}
\title{Cell-free Massive MIMO for UAV Communications}

\author{\IEEEauthorblockN{Carmen D'Andrea\IEEEauthorrefmark{1}, Adrian Garcia-Rodriguez\IEEEauthorrefmark{2}, Giovanni Geraci\IEEEauthorrefmark{3}, Lorenzo Galati Giordano\IEEEauthorrefmark{2}, and Stefano Buzzi\IEEEauthorrefmark{1}}
\IEEEauthorblockA{\IEEEauthorrefmark{1}\textit{University of Cassino and Southern Latium},
Cassino, Italy.}
\IEEEauthorblockA{\IEEEauthorrefmark{2}\textit{Nokia Bell Labs}, 
Dublin, Ireland.}
\IEEEauthorblockA{\IEEEauthorrefmark{3}\textit{Universitat Pompeu Fabra}, Barcelona, Spain.}
\thanks{\noindent Corresponding author: C. D'Andrea (carmen.dandrea@unicas.it).}
\thanks{\noindent The work of G. Geraci was partly supported by the Postdoctoral Junior Leader Fellowship Programme from ``la Caixa" Banking Foundation.}
}

\maketitle
\thispagestyle{empty}
\newcounter{MYtempeqncnt}

\begin{abstract}
We study support for unmanned aerial vehicle (UAV) communications through a cell-free massive MIMO architecture. Under the general assumption that the propagation channel between the mobile stations, either UAVs or ground users, and the access points follows a Ricean distribution, we derive closed form spectral efficiency lower bounds for uplink and downlink with linear minimum mean square error (LMMSE) channel estimation. We also propose power allocation and user scheduling strategies for such a system. Our numerical results reveal that a cell-free massive MIMO architecture may provide better performance than a traditional multicell massive MIMO network deployment. 
\end{abstract}
\IEEEpeerreviewmaketitle
\section{Introduction}
In the last few years, unmanned aerial vehicles (UAVs) have attracted a lot of attention, due to the availability of compact, small-size, energy-efficient models able to perform many critical tasks efficiently and in an automated manner. The integration of UAVs in wireless communication networks has become a hot research area, mainly with two different approaches \cite{MozSaaBen2018, DBLP:journals/corr/abs-1809-01752}. The first research approach focuses on the services that UAVs can bring to wireless networks, since UAVs can be regarded as moving access points (APs). With this perspective, UAVs can be used to increase the network capacity on-demand, fill network coverage holes, fastly deploy a mobile network architecture in the presence of a catastrophic event, etc. The second research approach focuses on the services that the network can bring to UAVs, and in particular on the use of a wireless network to support communications with UAVs \cite{LinYajMur2018,AzaRosPol2017,ZenLyuZha2018,LopDinLi2018GC}. Considering the latter approach, \cite{GarGerLop2018,GeraciUAVs_Access2018,GerGarGalICC2018} 
have recently investigated the use of massive MIMO (mMIMO) to support UAVs cellular communications, showing that equipping base stations (BSs) with large antenna arrays dramatically increases---with respect to a traditional cellular deployment---the probability of meeting the stringent reliability requirements of the UAVs command and control (C$\&$C) link.

In this paper, we investigate the use of cell-free (CF) and user-centric (UC) network deployments for the support of UAV communications. 
In a  CF massive MIMO architecture \cite{Ngo_CellFree2016}, large base stations with co-located massive MIMO arrays are replaced by a much larger number of APs, with a small number of antennas and reduced processing capabilities.  The APs are connected via a backhaul network to a central processing unit (CPU), which sends to the APs the data symbols to be transmitted to the users and receives soft estimates of the received data symbols from all APs. Neither channel estimates nor beamforming vectors are propagated through the backhaul network, and the time-division-duplex protocol is used to exploit uplink/downlink channel reciprocity.  The results in \cite{Ngo_CellFree2016} show that the CF approach  provides better performance than a small-cell system in terms of 95$\%$-likely per-user throughput. More recently,  \cite{buzziWCL2017, Buzzi_WSA2017} have introduced a user-centric (UC) virtual-cell massive MIMO approach to CF massive MIMO, assuming that each AP does not serve all the users in the system, but only a subset of them. The UC approach has been shown to provide better performance than the pure CF approach to the vast majority of the users in the network, since it allows APs to focus their available resources on the users that will benefit the most.

Following on such a track, in this paper we evaluate the capability of a CF UC massive MIMO deployment to support UAV  communications in the presence of legacy ground users (GUEs) using the same frequency band. Assuming a Ricean channel model, and linear minimum mean square error (LMMSE) channel estimation, the paper derives a lower bound to the achievable spectral efficiency for both the uplink and downlink.  The numerical results reveal the superiority of the considered architecture with respect to a traditional multicell massive MIMO network, for both UAVs and GUEs. 
\section{System model}

\subsection{Cell-Free Network Topology}
\begin{figure}[!t]
\centering
\includegraphics[width=\figwidth]{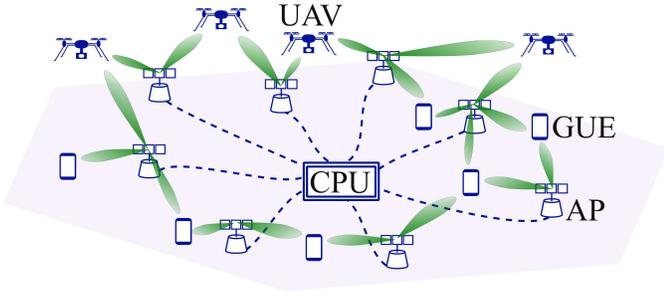}
\caption{Cell-free network supporting both ground and UAV users.}
\label{Fig:Ref_Sys}
\end{figure}

We consider a network that consists of outdoor APs, GUEs, and UAVs---as depicted in Fig.~\ref{Fig:Ref_Sys}---, whose sets are denoted by $\mathcal{A}$, $\mathcal{G}$, and $\mathcal{U}$, and have cardinalities $N_\mathrm{A}$, $N_\mathrm{G}$, and $N_\mathrm{U}$, respectively. In the following, we let the term \emph{users} denote both GUEs and UAVs. We assume that all users are equipped with a single antenna and that each AP is equipped with a uniform linear array (ULA) with $N_{\rm AP}$ antennas. We define $\mathcal{K} = \mathcal{G} \cup \mathcal{U}$, use $K=N_\mathrm{G}+N_\mathrm{U}$ to denote the number of users in the system, and let $\mathcal{K}_a$ with cardinality $K_a$ be the set of users served by the $a$-th AP on a given physical resource block (PRB). Moreover, we denote by $\mathcal{A}_k$ the set of APs serving the $k$-th user, with $A_k$ representing its cardinality.

The $N_\mathrm{A}$ APs are connected by means of a backhaul network to a CPU wherein data-decoding is performed. Building on the approach of \cite{Ngo_CellFree2016}, all communications take place on the same frequency band; uplink and downlink are separated through time-division-duplex (TDD). The coherence interval is thus divided into three phases: (a) uplink channel estimation, (b) downlink data transmission, and (c) uplink data transmission. 
In phase (a), users send pilot data in order to enable channel estimation at the APs. 
In phase (b), APs use channel estimates to perform channel-matched beamforming and send data symbols on the downlink. Finally, in phase (c), users send uplink data symbols to the APs.
Note that no pilots are transmitted on the downlink and no channel estimation is performed at the users.

\subsection{Propagation Channel}

We denote by $\mathbf{g}_{k,a} \in \mathbb{C}^{N_{\rm AP}}$ the channel between the $k$-th user and the $a$-th AP. We assume Ricean fading channel, which consist of a dominant line-of-sight (LOS) component on top of a Rayleigh-distributed component modelling the scattered multipath. The channel from the $k$-th user to the $a$-th AP is modelled as
\begin{equation}
\mathbf{g}_{k,a}=\sqrt{\frac{\beta_{k,a}}{K_{k,a}+1}} \left[ \sqrt{K_{k,a}} e^{j \vartheta_{k,a}} \mathbf{a}\left(\theta_{k,a}\right) + \mathbf{h}_{k,a}  \right] \; ,
\label{Channel_generic}
\end{equation}
where $\beta_{k,a}$ is a scalar coefficient modelling the channel path-loss and shadowing effects, $K_{k,a}$ is the Ricean $K-$factor, and the $\left(N_{\rm AP} \times 1\right)-$dimensional vector $\mathbf{h}_{k,a}$ contains the i.i.d. $\mathcal{CN} (0,1)$ small-scale fading coefficients between the $a$-th AP and the $k$-th user. Moreover, $\vartheta_{k,a}$ is a $\mathcal{U}[0, 2\pi]$ random variable representing a phase rotation. The vector $\mathbf{a}\left(\theta_{k,a}\right)$ is the AP antenna array steering vector corresponding to the  direction-of-arrival $\theta_{k,a}$. 
Letting $\mathbf{z}_{a,\ell}$ denote the vector containing the coordinates of the $\ell$-th antenna at the $a$-th AP, and denoting by $\widetilde{\mathbf{z}}_k$ the vector containing the coordinates of the $k$-th user, 
the $\ell$-th entry of the vector $\mathbf{a}\left(\theta_{k,a}\right)$ can be expressed as
\begin{equation}
\left[\mathbf{a}\left(\theta_{k,a}\right)\right]_{\ell}=e^{-j2\pi\left( \|\mathbf{z}_{a,1}-\widetilde{\mathbf{z}}_k \| - \|\mathbf{z}_{a,\ell}-\widetilde{\mathbf{z}}_k \|\right)}.
\end{equation}

In the following we further describe the above parameters depending on the specific link type:

\subsubsection{GUE-AP parameters} \label{GUE-AP_model}
With regard to the GUE-to-AP channel, i.e., when $k \in \mathcal{G}$, we assume that all the GUEs channels are Rayleigh-distributed, i.e., $K_{k,a}=0 , \; \forall a,k$. For the large scale coefficients $\beta_{k,a}$ we adopt the model of \cite{Ngo_CellFree2016}, i.e.
\begin{equation}
\beta_{k,a}= 10^{\frac{\text{PL}_{k,a}}{10}} 10^{\frac{\sigma_{\rm sh}z_{k,a}}{10}},
\label{Beta_GUEs}
\end{equation}
where $\text{PL}_{k,a}$ represents the path loss (expressed in dB) from the $k$-th GUE to the $a$-th AP, evaluated using the three-slope path loss model of \cite{Ngo_CellFree2016, buzziWCL2017}. Moreover, $10^{\frac{\sigma_{\rm sh}z_{k,a}}{10}}$ represents the shadow fading with standard deviation $\sigma_{\rm sh}$, where $z_{k,a}$ takes into account the correlation of the shadow fading between APs and GUEs that are in close proximity \cite{Ngo_CellFree2016, buzziWCL2017}.

\subsubsection{UAV-AP parameters} \label{UAV-AP_model}
With regard to the UAV-to-AP channel, i.e., when $k \in \mathcal{U}$,  the Ricean
factor $K$ is assumed to be a function of the UAV-AP distance \cite{Jafari_Lopez_2015}, i.e.
\begin{equation}
K_{k,a}=\frac{p_{\rm LOS}\left(d_{k,a}\right)}{1-p_{\rm LOS}\left(d_{k,a}\right)},
\end{equation}
where $d_{k,a}$ is the distance between the $k$-th user, and the $a$-th AP and $p_{\rm LOS}\left(d_{k,a}\right)$ is the LOS probability evaluated according to \cite[Table B-1]{3GPP_36777} for the UMi scenario. For the large scale fading we assume
\begin{equation}
\beta_{k,a}= 10^{\frac{\text{PL}_{k,a}}{10}},
\label{Beta_UAVs}
\end{equation}
with the path-loss evaluated according to \cite[Table B-2]{3GPP_36777} for the UMi scenario.

\subsection{User Association and Scheduling}
The set $\mathcal{K}_a$ of users associated to the $a$-th AP can be determined according to several criteria. In this paper, we consider the two following approaches.

\subsubsection{CF approach}
In the CF approach, each AP communicates with all the users in the system, i.e.  we have that $\mathcal{K}_a = \mathcal{K}, \; \forall \, a=1,\ldots,N_\mathrm{A}$ and the set $\mathcal{A}_k=\mathcal{A}, \; \forall \, k=1,\ldots,K$.

\subsubsection{UC approach}
In the UC approach, the $k$-th user is served by the $A_k$ APs that it receives with best average channel conditions. The set $\mathcal{A}_k$, contains the $A_k$ APs with the largest slow fading coefficients to the $k$-th user.

\section{The Communication Process} \label{Section_Comm}

\begin{figure*}[b!]
\hrulefill
\vspace*{1pt}
\setcounter{MYtempeqncnt}{\value{equation}}
\setcounter{equation}{15}
\begin{equation}
\begin{aligned}
\overline{\text{SINR}}_{k, {\rm DL}} &= \ds \left( \ds \sum_{a\in{\cal A}_k} {\ds \sqrt{\eta_{k,a}^{\rm DL}} \gamma_{k,a}} \right)^2 \times \Bigg\{
\ds \sum_{a\in{\cal A}_k} \eta_{k,a}^{\rm DL} \left( \eta_{k} \delta_{k,a}^{(k)}-\gamma_{k,a}^2\right)+ \ds \sum_{j \in \mathcal{K}} \sqrt{\eta_j} \ds \sum_{a\in{\cal A}_j} \eta_{j,a}^{\rm DL} \text{tr} \left( \mathbf{G}_{j,a} \mathbf{D}_{j,a}^H \mathbf{G}_{k,a} \right)  + \sigma^2_z \\
& \enspace + \ds \sum_{j \in \mathcal{K}\backslash k} \eta_k \bigg\{ \ds \sum_{a\in{\cal A}_j} \bigg[\eta_{j,a}^{\rm DL}\delta_{k,a}^{(j)} + \ds \sum_{\substack{b\in{\cal A}_j \\ b \neq a}} \sqrt{\eta_{j,a}^{\rm DL}} \sqrt{\eta_{j,b}^{\rm DL}} \text{tr} \left(\mathbf{D}_{j,a}\mathbf{G}_{k,a}\right)\text{tr} \left(\mathbf{D}_{j,b}^H\mathbf{G}_{k,b}\right)\bigg]\bigg\} \left|\boldsymbol{\phi}_k^H \boldsymbol{\phi}_j \right|^2 \Bigg\}^{-1}.
\end{aligned}
\label{eq:SINR_bar_DL}
\end{equation}
\setcounter{equation}{\value{MYtempeqncnt}}
\vspace*{-1cm}
\end{figure*}

\subsection{Uplink Training} \label{MMSE_Ch_est}
We denote by $\tau_c$ the length (in time-frequency samples) of the channel coherence time, and by $\tau_p$ the length (in time-frequency samples) of the uplink training phase, where we must ensure that $\tau_p < \tau_c$. 
Denote by $\boldsymbol{\phi}_k$ the $\tau_p$-dimensional column pilot sequence sent by the $k$-th user, and assume that $\|\boldsymbol{\phi}_k\|^2=1$, $ \forall \, k$.
The signal received at the $a$-th AP during the training phase can be expressed through the following $\left(N_{\rm AP} \times \tau_p \right)$-dimensional matrix
\begin{equation}
\mathbf{Y}_a = \ds \sum_{k \in \mathcal{K}} \ds \sqrt{\eta_k} \mathbf{g}_{k,a}\boldsymbol{\phi}_k^H + \mathbf{W}_a \; ,
\label{eq:y_m}
\end{equation}
with ${\eta}_k$ denoting the power employed by the $k$-th user during the training phase, and $\mathbf{W}_a$ a $\left(N_{\rm AP} \times \tau_p\right)$-dimensional matrix with i.i.d. ${\cal CN}(0, \sigma^2_w)$ entries containing the thermal noise contribution at the $a$-th AP.
Based on the observable $\mathbf{Y}_a$, and exploiting the knowledge of the users' pilot sequences, the $a$-th AP performs estimation of the channel vectors 
$\left\{\mathbf{g}_{k,a}\right\}_{k\in \mathcal{K}_a}$. We assume here knowledge of the user transmit powers $\left\{\eta_k\right\}_{k\in \mathcal{K}}$. 
Assuming knowledge of the large-scale fading coefficients $\beta_{k,a}$ as in \cite{Ngo_CellFree2016} and of the vectors $\mathbf{a}\left(\theta_{k,a}\right) \, \forall \; a,k$, we form a LMMSE estimate of $\left\{\mathbf{g}_{k,a}\right\}_{k\in \mathcal{K}_a}$ based on the $N_{\rm AP}$-dimensional statistics 
\vspace{-0.3cm}
\begin{equation}
\widehat{\mathbf{y}}_{k,a}=\mathbf{Y}_a \boldsymbol{\phi}_k= \sqrt{\eta_k}\mathbf{g}_{k,a} + \ds \sum_{\substack{i=1 \\ i\neq k}}^K {\sqrt{\eta_i}\mathbf{g}_{i,a}\boldsymbol{\phi}_i^H \boldsymbol{\phi}_k} + \mathbf{W}_a \boldsymbol{\phi}_k \; .
\label{y_hat_ka}
\end{equation}
The LMMSE channel estimate of the channel $\mathbf{g}_{k,a}$ is thus written as
$
\widehat{\mathbf{g}}_{k,a}= 
\mathbf{D}_{k,a} \,  \widehat{\mathbf{y}}_{k,a},
$
where the $\left( N_{\rm AP} \times N_{\rm AP}\right)-$dimensional matrix $\mathbf{D}_{k,a}$ can be written as
\begin{equation}
\mathbf{D}_{k,a, \rm{U}} = \sqrt{\eta_k} \mathbf{G}_{k,a} \mathbf{B}_{k,a}^{-1}\; ,
\label{D_ka}
\end{equation}
with
\begin{equation}
\mathbf{G}_{k,a} = \frac{\beta_{k,a}}{K_{k,a}+1} \left[ K_{k,a} \mathbf{a}\left(\theta_{k,a}\right)\mathbf{a}^H\left(\theta_{k,a}\right) +\mathbf{I}_{N_{\rm AP}}\right],
\label{G_ka}
\end{equation}
and
\begin{equation}
\mathbf{B}_{k,a} =\sum_{i \in \mathcal{K}} {\eta_i \beta_{i,a} \mathbf{G}_{i,a}\left|\boldsymbol{\phi}_i^H \boldsymbol{\phi}_k\right|^2 } + \sigma^2_w \mathbf{I}_{N_{\rm AP}}.
\label{B_ka}
\end{equation}

\subsection{Downlink Data Transmission} \label{DL_Section}

\begin{figure*}[b!]
\hrulefill
\vspace*{1pt}
\setcounter{MYtempeqncnt}{\value{equation}}
\setcounter{equation}{22}
\begin{equation}
\begin{aligned}
\overline{\text{SINR}}_{k, {\rm UL}} &= \ds \eta_{k}^{\rm UL}  \left( \ds \sum_{a\in{\cal A}_k} {\ds  \gamma_{k,a}} \right)^2 \times \Bigg\{ \ds \eta_{k}^{\rm UL} \sum_{a\in{\cal A}_k}  \left( \eta_{k} \widetilde{\delta}_{k,a}^{(k)}-\gamma_{k,a}^2\right)+ \ds \sum_{j \in \mathcal{K}} \eta_j^{\rm UL} \sqrt{\eta_k} \ds \sum_{a\in{\cal A}_k}  \text{tr} \left( \mathbf{G}_{k,a} \mathbf{D}_{k,a}^H \mathbf{G}_{j,a} \right)\\
 & \enspace + \sigma^2_w \sum_{a\in{\cal A}_k} {\ds  \gamma_{k,a}} + \ds \sum_{j \in \mathcal{K}\backslash k} \eta_j^{\rm UL} \eta_j \bigg\{ \ds \sum_{a\in{\cal A}_k} \bigg[\widetilde{\delta}_{j,a}^{(k)} + \ds \sum_{\substack{b\in{\cal A}_k \\ b \neq a}}  \text{tr} \left(\mathbf{D}_{k,a}^H\mathbf{G}_{j,a}\right)\text{tr} \left(\mathbf{D}_{k,b}\mathbf{G}_{j,b}\right)\bigg]\bigg\} \left|\boldsymbol{\phi}_j^H \boldsymbol{\phi}_k \right|^2 \Bigg\}^{-1}.
\end{aligned}
\label{eq:SINR_bar_UL}
\end{equation}
\setcounter{equation}{\value{MYtempeqncnt}}
\vspace*{-1cm}
\end{figure*}

The APs treat the channel estimates as the true channels and perform conjugate beamforming on the downlink. The signal transmitted by the $a$-th AP in a generic symbol interval is the following $N_{\rm AP}$-dimensional vector
\begin{equation}
\mathbf{s}_a = \ds \sum_{k\in{\cal K}_a}\ds \sqrt{\eta_{k,a}^{\rm DL}} \widehat{\mathbf{g}}_{k,a} {x}_k^{\rm DL} \; ,
\label{eq:transmittedscalar}
\end{equation}
with 
${x}_k^{\rm DL}$ the downlink data-symbol for the $k$-th user, and $\eta_{k,a}^{\rm DL}$ a scalar coefficient controlling the power transmitted by the $a$-th AP to the $k$-th user. Letting $\eta^{\rm DL}_a$ denote the overall transmitted power by the $a$-th AP, the normalized transmitted power must satisfy the constraint
\begin{equation}
\mathbb{E} \left[ \|\mathbf{s}_a \|^2\right]=\ds \sum_{k\in{\cal K}_a} {\eta_{k,a}^{\rm DL} \gamma_{k,a}} \leq \eta^{\rm DL}_a \; ,
\end{equation}
where 
$
\gamma_{k,a}=\mathbb{E}\left[ \widehat{\mathbf{g}}_{k,a}^H \widehat{\mathbf{g}}_{k,a}\right]=\sqrt{\eta_k}\text{tr} \left( \mathbf{G}_{k,a} \mathbf{D}_{k,a} \right)
$
, and $\text{tr} (\cdot)$ denotes the trace operator.

Subsequently, each user receives phase-aligned contributions from all APs and does not need to perform channel estimation. The generic $k$-th user receives the soft estimate for the data symbol
\begin{equation}
\begin{array}{llll}
\widehat{x}_k^{\rm DL} & =\ds \sum_{a \in \mathcal{A}}
\mathbf{g}_{k,a}^H \mathbf{s}_a + {z}_k  
 = \ds \sum_{a\in{\cal A}_k} \ds \sqrt{\eta_{k,a}^{\rm DL}} \mathbf{g}_{k,a}^H  \widehat{\mathbf{g}}_{k,a} {x}_k^{\rm DL} \\ & + 
\ds \sum_{j \in \mathcal{K}\backslash k} \ds \sum_{a \in \mathcal{A}_j} \sqrt{\eta_{j,a}^{\rm DL}}\mathbf{g}_{k,a}^H   \widehat{\mathbf{g}}_{j,a} {x}_j^{\rm DL} +  
 {z}_k\; ,
\end{array}
\label{eq:received_data_MS_UC}
\end{equation}
with ${z}_k$ being the ${\cal CN}(0, \sigma^2_z)$ additive white Gaussian noise (AWGN).

Given the expression in \eqref{eq:received_data_MS_UC} an upper bound (UB) for the achievable spectral efficiency can be obtained as \cite{Caire_bounds_2018}
\begin{equation}
\text{SE}_{k, {\rm UB}}^{\rm DL}\!=\!\ds\frac{\tau_{\rm d}}{\tau_c}\mathbb{E}\!\! \left[ \! 1+ \frac{\ds \left|\sum_{a\in{\cal A}_k}  \sqrt{\eta_{k,a}^{\rm DL}} \mathbf{g}_{k,a}^H  \widehat{\mathbf{g}}_{k,a}\right|^2}{\ds \sum_{j \in \mathcal{K}\backslash k} \left| \ds \sum_{a \in \mathcal{A}_j} \sqrt{\eta_{j,a}^{\rm DL}}\mathbf{g}_{k,a}^H   \widehat{\mathbf{g}}_{j,a}\right|^2\!\!\!\!+\!\sigma^2_z} \!\! \right] \, ,
\label{SE_DL_upper_bound}
\end{equation}
where $\tau_{\rm d}= \tau_c - \tau_p -\tau_{\rm u}$ and $\tau_{\rm u} $ are the lengths (in time-frequency samples) of  the downlink and uplink data transmission phases in each coherence interval, respectively. The expectation in \eqref{SE_DL_upper_bound} is made over the fast fading channel realizations.

In this paper we also derive a LB of the downlink spectral efficiency, $\text{SE}_{k,{\rm LB}}^{\rm DL}$. 
%

\emph{Lemma 1: } A LB of the downlink spectral efficiency is given by
\begin{equation}
\text{SE}_{k,{\rm LB}}^{\rm DL}= \ds\frac{\tau_{\rm d}}{\tau_c}  \log_2 \left( \ds 1 + \overline{\text{SINR}}_{k, {\rm DL}}\right) \; ,
\label{eq:SE_MMSE_LB}
\end{equation}
where $\overline{\text{SINR}}_{k, {\rm DL}}$ is shown in \eqref{eq:SINR_bar_DL} at the bottom of this page, and 

\setcounter{MYtempeqncnt}{\value{equation}}
\addtocounter{MYtempeqncnt}{1}
\setcounter{equation}{\value{MYtempeqncnt}}

\vspace{-0.3cm}
\begin{equation}
\begin{array}{llll}
&\delta_{k,a}^{(j)}=\ds \left( \frac{\beta_{k,a}}{K_{k,a}+1}\right)^2 \text{tr}^2\left(\mathbf{D}_{j,a}\right) \\& \ds + \left( \frac{\beta_{k,a}}{K_{k,a}+1}\right)^2 K_{k,a} \left[ \text{tr} \left( \mathbf{a}^H\left(\theta_{k,a}\right) \mathbf{D}_{j,a} \mathbf{a}\left(\theta_{k,a}\right) \mathbf{D}_{j,a}^H \right) \right.\\ & \left. + \text{tr} \left( \mathbf{a}^H\left(\theta_{k,a}\right) \mathbf{D}_{j,a}^H \mathbf{a}\left(\theta_{k,a}\right) \mathbf{D}_{j,a} \right)\right].
\end{array}
\label{delta_ka_j}
\end{equation}

\emph{Proof: } The proof of Lemma 1 is based on the application of the use-and-then-forget (UatF) bound \cite{marzetta2016fundamentals}. The details of the derivation are omitted due to lack of space. \hfill $\square$ 

\subsection{Uplink Data Transmission} \label{UL_Section}
Since users do not perform channel estimation, they just send their data symbols without any channel-dependent phase offset. The $N_{\rm AP}$-dimensional vector received at the $a$-th AP in a generic symbol interval is expressed as 
\begin{equation}
{\bar{\mathbf{y}}}_a=\ds \sum_{k \in \mathcal{K}} \ds \sqrt{\eta_{k}^{\rm UL}} \mathbf{g}_{k,a} {x}^{\rm UL}_k + \mathbf{w}_m \; ,
\end{equation}
with ${\eta_{k}^{\rm UL}}$ and ${x}^{\rm UL}_k$  representing the uplink transmit power and the data symbol of the $k$-th user, respectively, and $\mathbf{w}_m \sim {\cal CN}(\mathbf{0}, \sigma^2_w \mathbf{I} )$ is the $N_{\rm AP}$-dimensional AWGN vector.

Each AP decodes the data transmitted by users in ${\cal K}_a$. The $a$-th AP thus forms, for each $k \in {\cal K}_a$,  the  statistics
${{t}}_{a,k}= \widehat{\mathbf{g}}_{k,a}^H {\bar{\mathbf{y}}}_a$ and sends them to the CPU.  Accordingly, the CPU is able to perform the soft estimates for the data sent by the users as follows
\begin{equation}
\widehat{{x}}^{\rm UL}_k = \ds \sum_{a \in \mathcal{A}_k} {{t}}_{a,k} \; , \quad k \in \mathcal{K} \, .
\label{Est_UL_uc1_MS}
\end{equation}
%
Using straightforward manipulations, \eqref{Est_UL_uc1_MS} can be re-written as
\begin{equation}
\begin{array}{llll}
\widehat{x}_k^{\rm UL} &= \ds \sum_{a\in{\cal A}_k} \ds \sqrt{\eta_{k}^{\rm UL}} \widehat{\mathbf{g}}_{k,a}^H  \mathbf{g}_{k,a} {x}_k^{\rm UL} \\ & +
\ds \sum_{j \in \mathcal{K}\backslash k} \ds \sum_{a \in \mathcal{A}_k} \sqrt{\eta_{j}^{\rm UL}}\widehat{\mathbf{g}}_{k,a}^H   \mathbf{g}_{j,a} {x}_j^{\rm UL} +  
 \ds \sum_{a\in{\cal A}_k} {\widehat{\mathbf{g}}_{k,a}^H \mathbf{w}_a  } .
\end{array}
\label{Est_UL_uc1_MS2}
\end{equation}
Similarly to procedure followed for deriving \eqref{Est_UL_uc1_MS2}, an UB for the achievable spectral efficiency can be obtained as \cite{Caire_bounds_2018}
\vspace{-0.2cm}
\begin{equation}
\!\text{SE}_{k, {\rm UB}}^{\rm UL}\!=\!\ds\frac{\tau_{\rm u}}{\tau_c} \!  \mathbb{E} \!\!\left[ \! 1\!+\! \frac{\ds \eta_{k}^{\rm UL}\left|\sum_{a\in{\cal A}_k} \widehat{\mathbf{g}}_{k,a}^H  \mathbf{g}_{k,a}\right|^2}{\!\!\!\ds \sum_{j \in \mathcal{K}\backslash k} \!\eta_{j}^{\rm UL}\! \left| \ds \sum_{a \in \mathcal{A}_k} \widehat{\mathbf{g}}_{k,a}^H \mathbf{g}_{j,a}\right|^2\!\!\!+\!\sigma^2_w \!\!\!\!\sum_{a \in \mathcal{A}_k} \!\!\| \widehat{\mathbf{g}}_{k,a} \|^2}  \right].
\label{SE_UL_upper_bound}
\end{equation}

\emph{Lemma 2: } A LB for the uplink spectral efficiency can be expressed as
\begin{equation}
\text{SE}_{k,{\rm LB}}^{\rm UL}= \ds\frac{\tau_{\rm u}}{\tau_c}   \log_2 \left( \ds 1 + \overline{\text{SINR}}_{k, {\rm UL}}\right) \; ,
\label{eq:SE_MMSE_LB_UL}
\end{equation}
where $\overline{\text{SINR}}_{k, {\rm UL}}$ in \eqref{eq:SINR_bar_UL} is shown at the bottom of this page, and

\setcounter{MYtempeqncnt}{\value{equation}}
\addtocounter{MYtempeqncnt}{1}
\setcounter{equation}{\value{MYtempeqncnt}}

\vspace{-0.3cm}
\begin{equation}
\begin{array}{llll}
&\widetilde{\delta}_{j,a}^{(k)}=\ds \left( \frac{\beta_{j,a}}{K_{j,a}+1}\right)^2 \text{tr}^2\left(\mathbf{D}_{k,a}\right) \\& \ds + \left( \frac{\beta_{j,a}}{K_{j,a}+1}\right)^2 K_{j,a} \left[ \text{tr} \left( \mathbf{a}^H\left(\theta_{j,a}\right) \mathbf{D}_{k,a}^H \mathbf{a}\left(\theta_{j,a}\right) \mathbf{D}_{k,a} \right) \right. \\ & \left. + \text{tr} \left( \mathbf{a}^H\left(\theta_{j,a}\right) \mathbf{D}_{k,a} \mathbf{a}\left(\theta_{j,a}\right) \mathbf{D}_{k,a}^H \right)\right].
\end{array}
\label{delta_tilde_ka_j}
\end{equation}

\emph{Proof: } The details of the proof, which is also based on the UatF bound, are again omitted due to space constraints.  \hfill $\square$ 
\section{Power Allocation Strategies}

\subsection{Downlink Power Control}

\subsubsection{Proportional power allocation}
For the downlink data transmission, the first power allocation strategy that we consider is the proportional power allocation (PPA):
\begin{equation}
P_{k,a}^{\rm DL}=  \left\lbrace
\begin{array}{llll}
\ds \eta^{\rm DL}_a \frac{ \gamma_{k,a}}{\ds \sum_{j\in{\cal K}_a} {\gamma_{j,a}}}, \; & \text{if} \; k \in \mathcal{K}_a
\\
0 & \text{otherwise}
\end{array} \right. ,
\end{equation}
where $P_{k,a}^{\rm DL}=\eta_{k,a}^{\rm DL} \gamma_{k,a}$ is the power transmitted by the $a$-th AP to the $k$-th user. This power allocation rule is such that the generic $a$-th AP divides its power $\eta_a^{\rm DL}$ in a way that is proportional to the estimated channel strengths. This way, users with good channel coefficients will receive a larger share of the transmit power than users with bad channels.

\subsubsection{Waterfilling power control}
The second power allocation that we consider is a modified waterfilling power control (WFPC), where the ``noise'' level for the communication between the $a$-th AP and the $k$-th user is written as 
\begin{equation}
L_{k,a}= \frac{\sigma^2_z}{\gamma_{k,a}}.
\end{equation}
The WFPC gives the following power allocation
\begin{equation}
P_{k,a}^{\rm DL}=  \left\lbrace
\begin{array}{llll}
\left( \nu_a - L_{k,a} \right)^+, \; & \text{if} \; k \in \mathcal{K}_a
\\
0 & \text{otherwise}
\end{array} \right. ,
\end{equation}
where $\nu_a$ is the water level and $(\cdot)^+$ denotes the positive part operator, with the constraint 

\begin{equation}
\sum_{k \in \mathcal{K}_a}\left( \nu_a - L_{k,a} \right)^+ = \eta^{\rm DL}_a \, .
\end{equation}
This heuristic power allocation rule can be seen as a sort of AP-centric approach to the CF massive MIMO system and is based on the well-known waterfilling algorithm \cite{Cover-Thomas}, which allocates a larger amount of power to the users with better channels conditions, i.e., to those with the lower ``noise'' levels.

\subsection{Uplink Power Control}

For the uplink data transmission, we consider standard fractional power control (FPC) \cite{BarGalGar2018GC, 3GPP_36777}, where the $k$-th user transmit power is given by 

\vspace{-0.4cm}
\begin{equation}
\eta_k^{\rm UL}= \text{min} \left( P_{\rm max}^{\rm UL}, P_0 \zeta_k^{-\alpha}\right) \, ,
\end{equation}
and the parameter $\zeta_k$ is obtained considering the channels from the $k$-th user to all the APs in the set $\mathcal{A}_k$ as

\begin{equation}
\zeta_k= \sqrt{\sum_{a \in \mathcal{A}_k} {\text{tr}^2\left( \mathbf{G}_{k,a} \right)}}.
\end{equation}
\section{Numerical Results and Key Insights}

\begin{table}
\centering
\caption{Cell-free system parameters}
\label{table:parameters}
\def\arraystretch{1.2}
\begin{tabulary}{\columnwidth}{ |p{2.8cm}|p{4.7cm}| }
\hline
	\textbf{Deployment} 			&  \\ \hline
  	AP distribution				&  Horizontal: uniform, vertical: 15~m \\ \hline
  GUE distribution 				& Horizontal: uniform, vertical: 1.65~m\\ \hline
	UAV distribution 				& Horizontal: uniform, vertical uniform between 22.5~m and 300~m \cite{3GPP_36777}\\ \hline \hline
	\textbf{PHY and MAC} 			&  \\ \hline
	Carrier freq., bandwidth		&  $f_0=1.9$ GHz, $W = 20$ MHz \\ \hline
	AP antenna array			& Four-element ULA with $\lambda/2$ spacing\\ \hline
	User antennas 		& Omnidirectional with 0~dBi gain\\ \hline
	\multirow{2}{*}{Power control}		& DL: proportional power allocation (PPA) or waterfilling power control (WFPC) \\ \cline{2-2}
	& UL: FPC with $\alpha=0.5$, $P_0=-35$ dBm \\ \hline
	Thermal noise 				& -174 dBm/Hz spectral density \\ \hline
	Noise figure 			& 9 dB at APs/GUEs/UAVs \\ \hline
	User association		& Cell free (CF) or user centric (UC) \\ \hline
	Traffic model		& Full buffer \\ \hline
\end{tabulary}
\end{table}

We consider a square area of 1 km$^2$ with $N_\mathrm{A}=100$ APs, $N_{\rm G}=48$ GUEs and $N_{\rm U}=12$ UAVs. To avoid boundary effects, and to emulate a network with an infinite area, the square area is wrapped around at the edges. We assume $\tau_p=32$ that and that orthogonal pilots are randomly assigned to the users in the system, i.e., our results account for the impact of pilot contamination. The uplink transmit power during training is $\eta_k=\tau_p \overline{\eta}_k$, with $\overline{\eta}_k=100$~mW $ \forall k \in \mathcal{K}$. Regarding power allocation, we assume that the maximum downlink power transmitted by the $a$-th AP is $\eta^{\rm DL}_a= 200$~mW, $\forall a \in \mathcal{A}$, and the maximum uplink power transmitted by the $k$-th user is $P_{\rm max}^{\rm UL}=100$ mW, $\forall k \in \mathcal{K}$.  We consider $\tau_c = 200$ samples, corresponding to a coherence bandwidth of $200$ kHz and a coherence time of $1$~ms \cite{Ngo_CellFree2016}, and $\tau_{\rm d}=\tau_{\rm u}=\frac{\tau_c-\tau_p}{2}$. The remaining system parameters are detailed in Table \ref{table:parameters}.  
In the following, we report the rate per user, obtained as the product of the spectral efficiency---as per Section \ref{Section_Comm}---and the system bandwidth $W$. We also show the benchmark performance in the case of (i) perfect channel state information (PCSI), and (ii) a multicell massive MIMO (mMIMO) system with four 100-antenna BSs transmitting 8 W each. 

 
\subsubsection{Uplink performance}
\begin{figure}[!t]
\centering
\includegraphics[width=\figwidth]{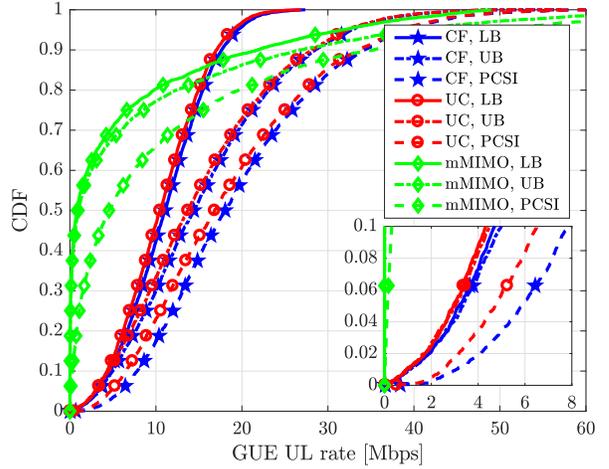}
\caption{UL rates for GUEs under: (i) cell-free (CF), (ii) user-centric (UC) with $A_k=10$, and (iii) multicell mMIMO (mMIMO) approaches.}
\label{Fig:GUE_UL}
\end{figure}

\begin{figure}[!t]
\centering
\includegraphics[width=\figwidth]{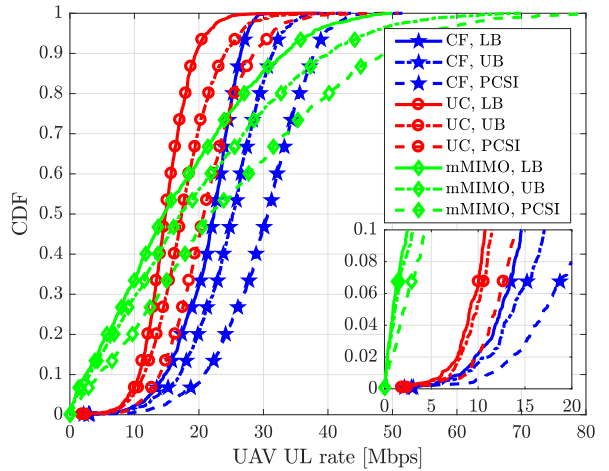}
\caption{UL rates for UAVs under: (i) cell-free (CF), (ii) user-centric (UC) with $A_k=10$, and (iii) multicell mMIMO (mMIMO) approaches.}
\label{Fig:UAV_UL}
\end{figure}

Figs. \ref{Fig:GUE_UL} and \ref{Fig:UAV_UL} report the cumulative distribution functions (CDFs) of the uplink (UL) rate for GUEs and UAVs, respectively, under: (i) a cell-free architecture (CF), (ii) a user-centric architecture (UC)---both under FPC---, and (iii) a benchmark multicell mMIMO deployment. Both figures show the advantages granted by the use of CF and UC schemes with respect to a classical multicell mMIMO deployment. In particular:
\begin{itemize}[leftmargin=*]
\item Due to the UL interference caused by UAVs---each in LOS with multiple BSs---the rate of many GUEs under a multicell mMIMO setup is limited. The percentage of GUEs in outage is reduced under perfect CSI, but the overall performance still remains negatively affected by the residual UAV-to-BS interference.
\item A distributed network architecture significantly improves the rates of the most vulnerable GUEs by bringing the APs in close proximity with them. Similar gains are achieved under CF and UC approaches, and they amount to over one order of magnitude for the 95$\%$-likely rate. Only the best GUEs, which happen to be located close to their serving BS, are better off under a multicell mMIMO setup.
\item While for UAVs the baseline performance of multicell mMIMO is not as bad as it is for GUEs, similar observations can be made. The most vulnerable UAVs strongly benefit from a distributed architecture that turns UAV-to-BS interference into useful signal. Additionally, a CF approach is preferable to UC, since UAVs experience good LOS propagation conditions with a large number of APs, and thus benefit from having many---rather than a subset of---APs serving them.
\end{itemize}

\subsubsection{Downlink performance}
\begin{figure}[!t]
\centering
\includegraphics[width=\figwidth]{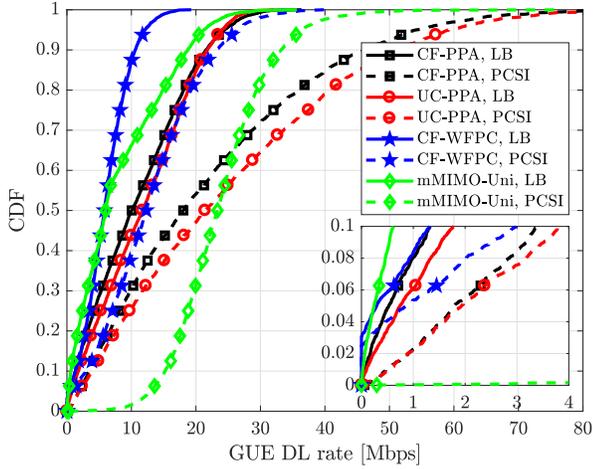}
\caption{DL rates for GUEs under: (i) cell-free with proportional power allocation (CF-PPA), (ii) user-centric with $A_k=10$ and proportional power allocation (UC-PPA), (iii) cell-free with waterfilling power control (CF-WFPC), and (iv) multicell mMIMO with uniform power (mMIMO-Uni).}
\label{Fig:GUE_DL}
\end{figure}

\begin{figure}[!t]
\centering
\includegraphics[width=\figwidth]{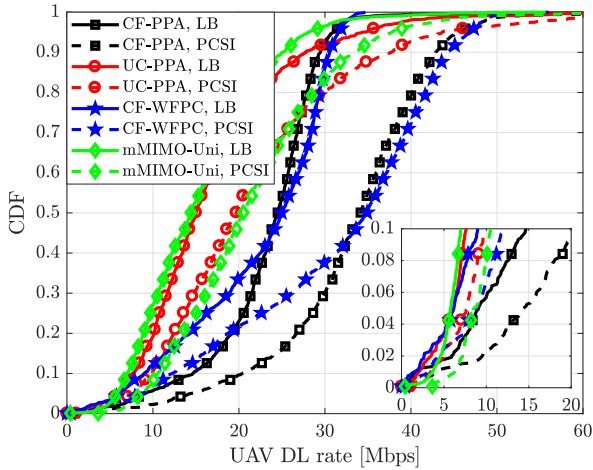}
\caption{DL rates for UAVs under: (i) cell-free with proportional power allocation (CF-PPA), (ii) user-centric with $A_k=10$ and proportional power allocation (UC-PPA), (iii) cell-free with waterfilling power control (CF-WFPC), and (iv) multicell mMIMO with uniform power (mMIMO-Uni).}
\label{Fig:UAV_DL}
\end{figure}

Figs. \ref{Fig:GUE_DL} and \ref{Fig:UAV_DL} show the CDFs of the downlink (DL) rate for GUEs and UAVs, respectively, for the following architectures: (i) cell-free with proportional power allocation (CF-PPA), (ii) user-centric with $A_k=10$ and proportional power allocation (UC-PPA), (iii) cell-free with waterfilling power control (CF-WFPC), and (iv) multicell mMIMO with uniform power allocation (mMIMO-Uni). Based on these figures, the following observations can be made:
\begin{itemize}[leftmargin=*]
\item The DL GUE performance under multicell mMIMO is affected by pilot contamination caused by the UAVs in the UL channel estimation phase. This is illustrated by the gap between the lower bound (LB) and the rates obtained under perfect CSI (PCSI). A user-centric approach with a fair power allocation policy (CF-PPA and UC-PPA) provides substantial gains.
\item A cell-free network brings significant benefits to the UAV DL---particularly under WFPC---owed to a large number of APs that serve each UAV and thus generate useful signal from what would otherwise be inter-cell interference.
\item A greedy waterfilling power control (WFPC) favors UAVs over GUEs, since UAVs end up being allocated more power due to their better channel conditions.
\end{itemize}

\section{Conclusions}

In this paper, we have investigated the use of cell-free and user-centric architectures for supporting wireless communications with UAVs. From the  derived spectral efficiency bounds, we have demonstrated that user-centric and cell-free network deployments can outperform multicell mMIMO networks, and that the improvements are particularly noticeable for the users with worst performance. An extension of this study will consider the use of more sophisticated power control rules, introducing strict reliability requirements for UAV communications, as well as the derivation of spectral efficiency formulas suited for the finite blocklength regime.

\ifCLASSOPTIONcaptionsoff
  \newpage
\fi
\bibliographystyle{IEEEtran}
\bibliography{Strings_Gio,Bib_Gio,References}

\end{document}